\documentclass[%
reprint,
superscriptaddress,
amsmath,amssymb,
aps,
prb,
]{revtex4-2}
\usepackage{textcomp}
\usepackage{graphicx}
\usepackage{dcolumn}
\usepackage{bm}
\usepackage{hyperref}

\begin{document}

\title{Development of a Capacitance Measurement for Pulsed Magnetic Fields}

\author{William K.~Peria}
\email{william.peria@nist.gov}
\altaffiliation{Present Address: National Institute of Standards and Technology, Boulder, Colorado 80305, USA}
\affiliation{National High Magnetic Field Laboratory, Los Alamos National Laboratory, Los Alamos, New Mexico 87545, USA}%

\author{Shengzhi Zhang}
\affiliation{National High Magnetic Field Laboratory, Los Alamos National Laboratory, Los Alamos, New Mexico 87545, USA}%

\author{Sangyun Lee}
\affiliation{National High Magnetic Field Laboratory, Los Alamos National Laboratory, Los Alamos, New Mexico 87545, USA}

\author{Vivien S.~Zapf}
\affiliation{National High Magnetic Field Laboratory, Los Alamos National Laboratory, Los Alamos, New Mexico 87545, USA}

\author{Choongjae Won}
\affiliation{Laboratory for Pohang Emergent Materials and Max Planck POSTECH Center for Complex Phase Materials, Department of Physics, Pohang University of Science and Technology, Pohang 37673, Korea}

\author{Sang-Wook Cheong}
\affiliation{Laboratory for Pohang Emergent Materials and Max Planck POSTECH Center for Complex Phase Materials, Department of Physics, Pohang University of Science and Technology, Pohang 37673, Korea}
\affiliation{Keck Center for Quantum Magnetism and Department of Physics and Astronomy, Rutgers University, Piscataway, New Jersey 08854, USA}

\author{Minseong Lee}
\email{ml10k@lanl.gov}
\affiliation{National High Magnetic Field Laboratory, Los Alamos National Laboratory, Los Alamos, New Mexico 87545, USA}%

\date{\today}

\begin{abstract}
Capacitance measurements are crucial for probing the electrical properties of materials. In this study, we develop and implement a capacitance measurement technique optimized for pulsed magnetic fields. Our approach employs an auto-balancing bridge method, leveraging a high-bandwidth transimpedance amplifier to mitigate parasitic contributions from coaxial cables. This technique enables precise capacitance measurements in rapidly changing magnetic fields, as demonstrated in experiments on the magnetoelectric material NiCo$_{2}$TeO$_{6}$. The results reveal strong magnetoelectric coupling, including a pronounced hysteresis in capacitance that coincides with magnetization measurements and an enhanced energy dissipation peak at high sweep rates. Compared to traditional LCR meter measurements in DC fields, our method exhibits excellent agreement while providing additional insight into field-induced phase transitions. This work establishes a robust methodology for capacitance measurements in extreme conditions and opens new opportunities for studying multiferroic and correlated electron systems under high magnetic fields.
\end{abstract}

\maketitle


\section{\label{sec:intro}Introduction}
Capacitance measurements are a foundational aspect of modern scientific research and material studies, playing a critical role in deepening our understanding of both natural and engineered materials. The capacitance of the materials provides invaluable insights into the fundamental material properties, such as the dielectric constant and loss factor, which are key to the development of innovative materials and devices. Furthermore, the capacitance measurement technique can also be extended to other measurements, for example, capacitive thermometry \cite{xia2007kapton,murphy2001capacitance}, capacitive torque magnetometry \cite{anand2022investigation}, and capacitive dilatometry \cite{martien2018ultrasensitive}, to name a few. For these reasons, the capacitance measurement is employed by research and engineering laboratories across the world.


In the study of material properties, researchers investigate the response of materials by varying external parameters. Common external parameters include temperature, pressure, and magnetic fields. Among these, magnetic fields couple to magnetic materials rather weakly, requiring high magnetic fields to control the properties of materials \cite{wampler2024magnetoelectric,blockmon2024high,lee2023field,hughey2022high,zhang2023electronic}. Following the research trend of demanding increasingly strong magnetic fields, efforts in high magnetic field studies are focused on developing magnets capable of generating even higher fields \cite{rubin1984high, motokawa2004physics,national2024current}.

\begin{figure}
\includegraphics[width=1\columnwidth]{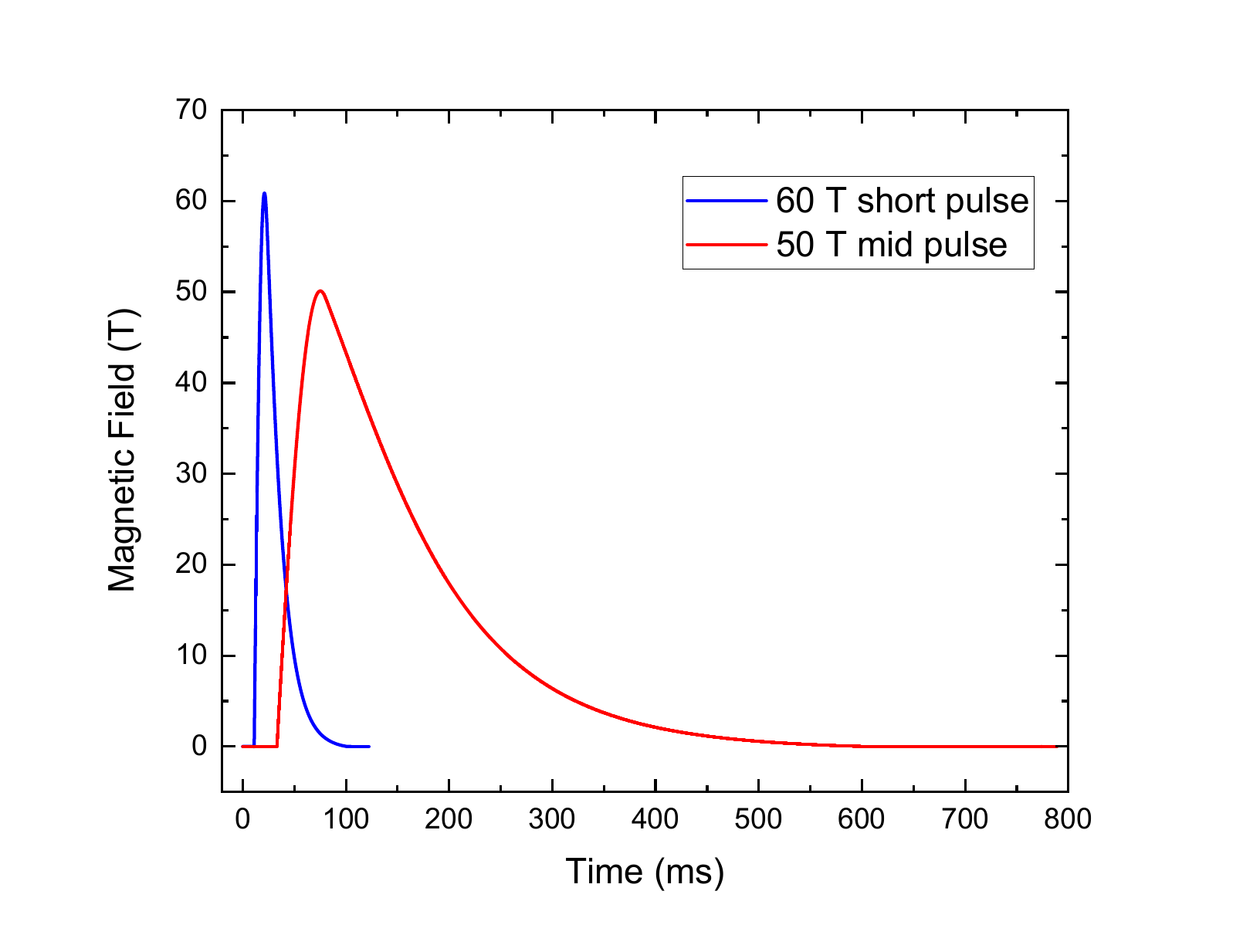}
\caption{\label{fig:overview} Time-dependent magnetic field profiles of the short-pulse and mid-pulse magnets at the National High Magnetic Field Laboratory, Los Alamos National Laboratory.}
\end{figure}

One method to achieve high magnetic fields is the use of pulsed magnetic fields \cite{Fritz_Herlach_1999}. Simply put, pulsed magnetic fields are generated by releasing the energy stored in a capacitor bank into a solenoid electromagnet over a very short period, creating an extremely high magnetic field. For instance, the 65~T short pulse magnet most commonly used at the National High Magnetic Field Laboratory located at Los Alamos National Laboratory reaches its maximum field from 0~T in 8.5~ms and then rapidly decays back to 0~T over 100~ms. A slower mid-pulse magnet can reach up to 59~T, operating approximately five to six times more slowly than the short pulse magnet [see Fig.~\ref{fig:overview}(a)]. The details of the magnets can be found in \cite{nguyen2016status,michel2020design,michel2023design}. 

In such ultra-fast environments, commercially available capacitance bridges or LCR meters cannot be used. This limitation arises because their measurement speeds are restricted to several milliseconds due to the time required for bridge balancing and data averaging, which makes them clearly incompatible with the timescales of the pulsed fields described above. As a result, these attempts are very rare. One method utilizes a commercial General Radio (GR) capacitance bridge, where the bridge is balanced prior to the measurement, and the unbalanced signal is then used for the measurement \cite{miyake2020capacitive}. While this allows for very accurate capacitance measurements, there are several drawbacks. Not only has the production of GR bridges been discontinued, but their maximum operating frequency is limited to 100~kHz, making it difficult to measure capacitance at higher frequencies. Additionally, this limitation of the frequencies makes it difficult to experiment with relatively faster short pulse magnets. Lastly, if a phase transition occurs and the capacitance is divergent, the unbalanced signal deviates from a linear relationship making precise quantification requiring additional data processes. 

In this paper, we employ the so-called auto-balancing bridge method, which leverages the high-quality virtual ground at the input of a transimpedance amplifier to measure capacitance in rapidly changing environments using transport measurement techniques. The primary challenge in capacitance measurements via transport techniques arises from the coaxial cable, which is connected in parallel with the sample. Consequently, the measured capacitance includes contributions from both the sample and the coaxial cable. While the sample capacitance is typically only a few picofarads (pF), a 1-meter coaxial cable can contribute several hundred pF, making it challenging to extract the sample signal accurately. As detailed later, introducing a virtual ground to both the sample and the measurement coaxial cable effectively separates the sample signal from the coaxial contribution, enabling precise detection of the AC current flowing through the sample.

\section{\label{sec:circuitry}Capacitance measurements in cryostats and pulsed-field magnets}
\begin{figure}
\includegraphics[width=1\columnwidth]{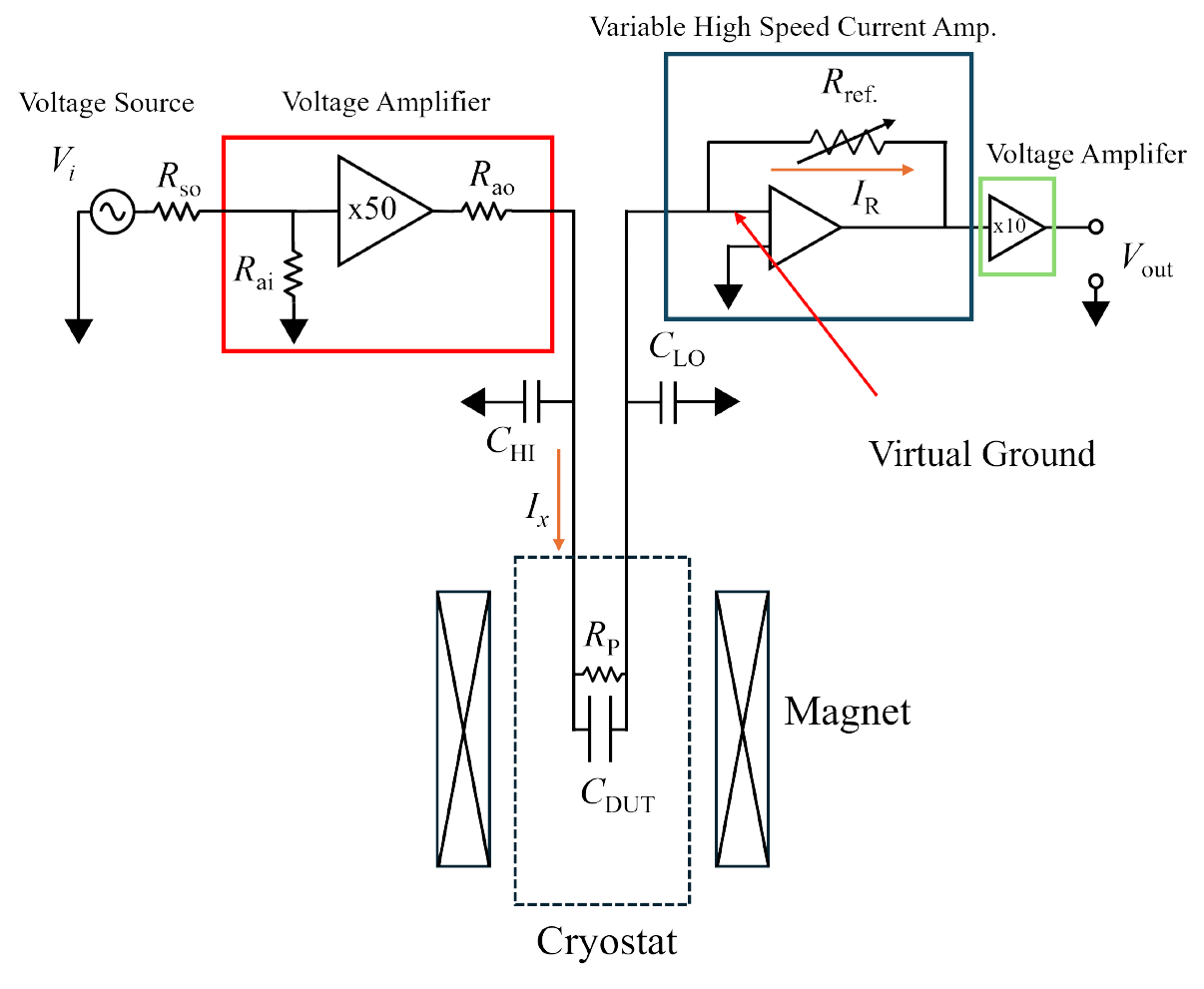}
\caption{\label{fig:circuit} An overview of the circuit for the capacitance measurement system in a pulsed magnetic field presented in this paper.}
\end{figure}
To measure the capacitance of the device under test (DUT) at low temperatures and in high magnetic fields, we employ a custom-built probe with the sample positioned at its end. Shielded coaxial cables are used for both excitation voltage and signal detection to minimize external noise. The residual capacitance of these cables is represented in Fig.~\ref{fig:circuit} as $C_{\text{HI}}$ for the cable connecting the excitation voltage to the DUT and $C_{\text{LO}}$ for the cable linking the DUT to the signal measurement station. The capacitance of these coaxial cables typically ranges from a few tens to several hundred pF, which can be up to two orders of magnitude larger than that of many DUTs.

\subsection{Oscillating Voltage Source}
\begin{figure}
\includegraphics[width=1\columnwidth]{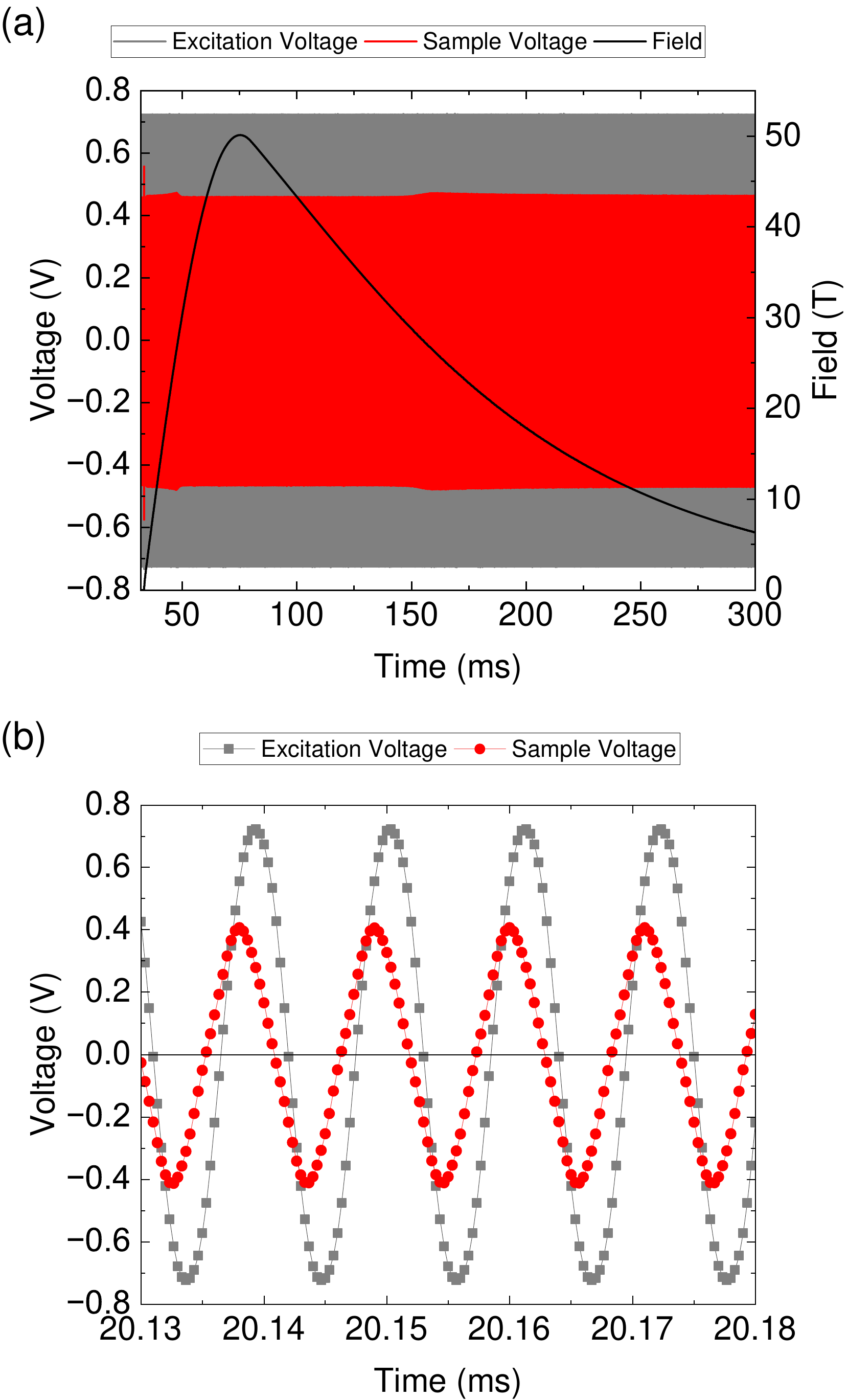}
\caption{\label{fig:rawdata} (a) Excitation voltage and sample signals as functions of time, recorded by the digitizer, along with the magnetic field strength as a function of time. (b) Enlarged view of the pre-trigger region from (a).}
\end{figure}

An analog AC voltage with a sinusoidal waveform $V_{in}$, generated by a function generator with a negligible output impedance, operates at 101~kHz with an RMS amplitude of 0.6~V [Fig.~\ref{fig:rawdata}(a)]. This signal is fed into a commercially available voltage amplifier (Advanced Energy, Trek 2100HF), which has an input impedance of $R_{\text{ai}}$ of 50~$\Omega$. The signal is amplified by a factor of 50 to achieve a high signal-to-noise (S/N) ratio. 

The excitation frequency is chosen to be sufficiently high to ensure an adequate number of data points are collected within the rise time of the field pulse while remaining below the capacitor’s self-resonant frequency to prevent the parasitic inductance from dominating the signal. However, it can be adjusted as needed. Since the amplifier’s output with a small output impedance $R_{ao}$ is directly connected to one side of the DUT, the effect of $C_{\text{HI}}$ can be safely neglected.

\subsection{Auto-Balancing Bridge}
A commercial high-bandwidth transimpedance amplifier (TIA) (Variable Gain High Speed Transimpedance Amplifier (Current Amplifier) DHPCA-100, FEMTO$^{\text{\textregistered}}$) is directly connected to the DUT, with its gain optimized based on the excitation frequency, typically around $10^3$. The TIA’s input is held at a virtual ground, ensuring that the current through the sample ($I_x$ in Fig.~\ref{fig:circuit}) equals the current through the range resistor ($I_R$ in Fig.~\ref{fig:circuit}), thereby eliminating parasitic contributions from coaxial cables $C_{\text{LO}}$ in the capacitance measurement \cite{ABB}. The TIA output is fed to a commercially available voltage amplifier (SR 560, Stanford Research Systems) to increase the S/N ratio by amplifying and filtering the signal. 

\subsection{Voltage detection}
The output voltage of the amplifier $V_{out}$ and the input voltage $V_{in}$ are recorded simultaneously using a high-bandwidth digitizer from National Instruments. These measured values are then used to calculate the complex impedance of the DUT, $Z_{tot}$, according to
\begin{align}
    Z_{tot}^{-1} &= \frac{1}{R_{ref}}\frac{V_{out}}{V_{in}} \\
                 &= |Z_{tot}^{-1}|\exp{(i\phi)}~\text{.}
\end{align}
In the frequency range of interest, the impedance of the DUT is on the order of 1~M$\Omega$. Therefore, it is appropriate to determine the capacitance of the sample by modeling it as a parallel resistor-capacitor circuit:
\begin{equation}
 Z_{tot}^{-1} = i\omega C_{DUT} + \left(\frac{1}{R_{P}}\right)~\text{.}
\end{equation} 

By solving for the capacitance and resistance, we obtain the following:

\begin{align}
    C_{DUT} &= \left|\frac{Z_{tot}^{-1}}{\omega}\right |\sin{\phi}\\
    R_{P} &= \frac{1}{|Z_{tot}^{-1}|\cos{\phi}}~\text{.}
    \label{eq:C}
\end{align}
The dissipation factor is commonly expressed as
\begin{equation}
    \tan(\delta) = \frac{1}{\omega R_{P}C_{DUT}}~\text{.}
    \label{eq:tan}
\end{equation} 

\section{\label{sec:results}Results}
\subsection{\label{sec:sample}Device under test---a multiferroic and magnetoelectric material}
A suitable DUT for our experiment is a magnetoelectric material, where the electric and magnetic properties are strongly coupled. In this system, an external magnetic field change the electric properties reflected as a change in capacitance through the change both from dielectric constant \cite{hur2004electric,sparks2014magnetocapacitance} and dimensional change \cite{ding2018measurement,jaime2017fiber}. 

In this report, we measured a high quality single crystal of NiCo$_{2}$TeO$_{6}$ \cite{skiadopoulou2020structural,won2025multi}. NiCo$_{2}$TeO$_{6}$ crystallizes in the noncentrosymmetric rhombohedral $R3$ structure and exhibits an incommensurate helical antiferromagnetic order with spins confined to the $ab$-plane. Its magnetic ordering occurs at around 52~K, and its magnetic propagation vector is determined as $k$ = (0, 0, 1.2109(1)). Notably, NiCo$_{2}$TeO$_{6}$ undergoes a hysteretic spin-flop transition at an external magnetic field of approximately 4~T. Dielectric anomalies coincide with its antiferromagnetic phase transition, and comprehensive terahertz and Raman spectroscopic studies reveal several low-frequency spin excitations---up to six distinct magnons at 5~K---that are highly sensitive to external magnetic fields, including the emergence of new modes and splitting near the spin-flop transition, which strongly indicate the presence of strong magnetoelectric coupling and electrically-active electromagnons \cite{pimenov2006possible}.

\subsection{Experimental data}
In this measurement, the excitation voltage frequency is 101.01~kHz, and the sampling rate is 3.3333~MHz. The phase difference between the excitation voltage and the sample signal is clearly visible, primarily arising from the voltage amplifiers and the impedance of the DUT.

\subsection{Capacitance versus Field}
\begin{figure}
\includegraphics[width=1\columnwidth]{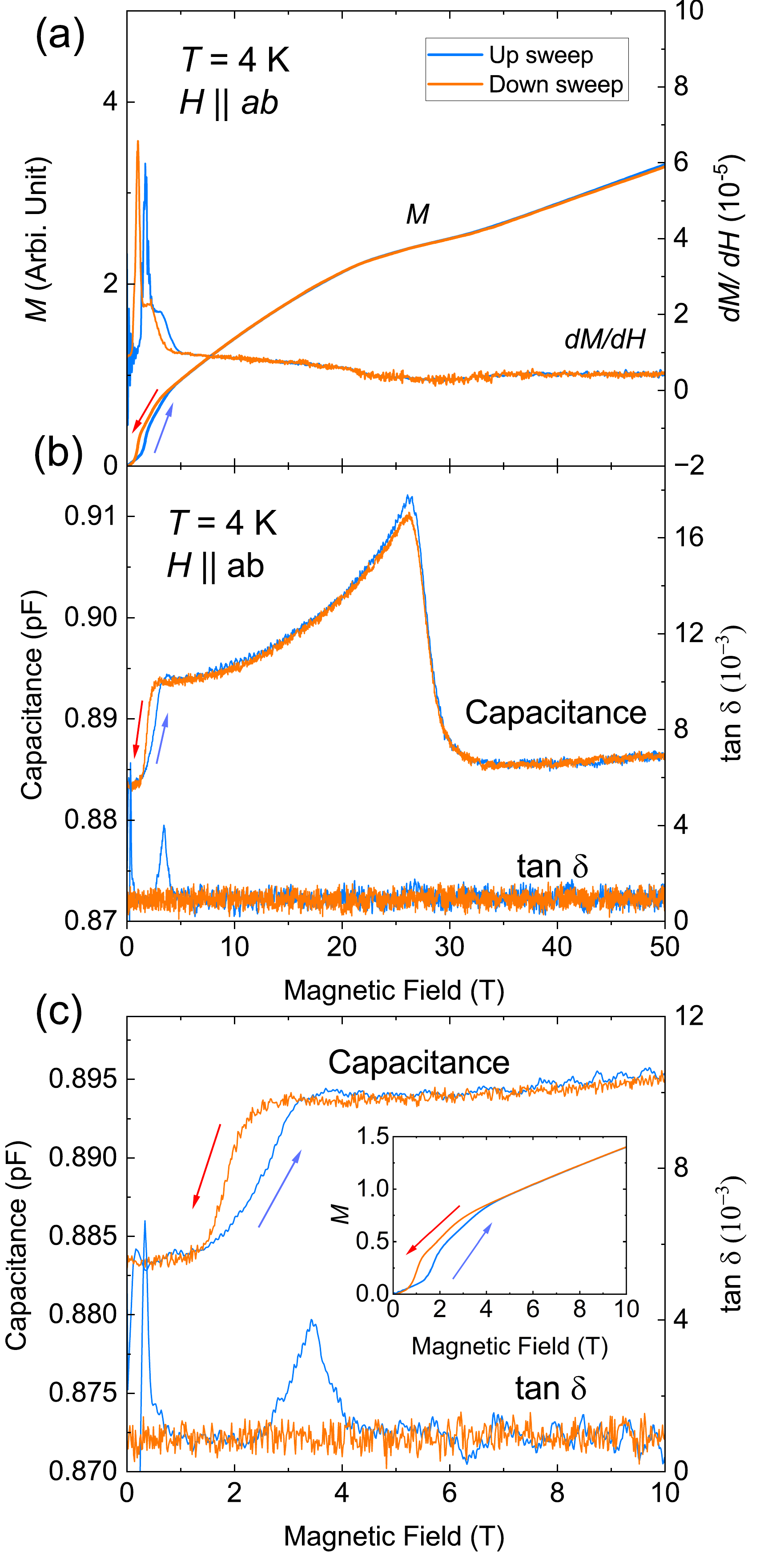}
\caption{\label{fig:MCt} (a) Magnetic field dependence of magnetization and its susceptibility ($dM/dH$) for NiCo$_{2}$TeO$_{6}$ with the field applied along the $ab$-plane at 4~K. (b) Magnetic field dependence of capacitance and tan($\delta$), measured with the magnetic field along the $ab$-plane and the electric field perpendicular to the $ab$-plane, from 0~T to 50~T at 4~K. (c) Enlarged view of the capacitance, tan($\delta$), and magnetization in the field range of 0~T to 10~T.}
\end{figure}
Figure \ref{fig:MCt} presents the magnetization, capacitance [Eq.~(\ref{eq:C})], and tan($\delta$) [Eq.~(\ref{eq:tan})] of NiCo$_{2}$TeO$_{6}$ as a function of the magnetic field applied along the $ab$-plane. The oscillatory voltage excitation applies an electric field perpendicular to the $ab$-plane. These measurements were performed at 4~K over a magnetic field range from 0~T to 50~T using the mid-pulse magnet.

In principle, the measured capacitance could be converted into the dielectric constant by assuming an infinite parallel plate capacitor geometry, given that the crystal is relatively thin compared to its lateral dimensions. The dielectric constant $\epsilon$ would then be given by
\begin{equation}
    \epsilon = \frac{Cd}{A},
\end{equation}
where $C$ is the capacitance, $d$ is the thickness of the crystal, and $A$ is the electrode area. However, we opted not to perform this conversion, as it relies on the assumption that both the thickness and area remain unchanged throughout the measurement. In a high magnetic field, mechanical deformations or magnetostriction effects could alter these dimensions, introducing uncertainties in the extracted dielectric constant. To rigorously determine $\epsilon$, one would need to independently measure the magnetic field dependence of both $d$ and $A$, ensuring that any potential changes are properly accounted for in the calculation.

As shown in Fig.~\ref{fig:MCt}~(a), the magnetization exhibits a clear hysteresis loop between 0 and 4~T, which is more prominently highlighted in the inset of Fig.~\ref{fig:MCt}~(c). Due to the steep changes in magnetization within this hysteresis region, the magnetic susceptibility ($dM/dH$) displays pronounced peaks, as seen in Fig.~\ref{fig:MCt}~(a). Our capacitance measurements shown in Fig.~\ref{fig:MCt}~(b) and Fig.~\ref{fig:MCt}~(c) reveal a corresponding hysteresis near the same magnetic fields, indicating a strong magnetoelectric coupling in this compound. 
Interestingly, the tan($\delta$) data exhibit a pronounced peak during the upsweep of the magnetic field, suggesting an energy dissipation process associated with the hysteresis loop. Another notable feature, albeit weaker, appears around 25~T as a subtle change in the slope of the magnetization curve. Remarkably, this weak change in magnetization coincides with a significant variation in capacitance, underscoring the sensitivity of capacitance measurements in detecting subtle magnetic state changes. This observation directly suggests that the weak slope change in magnetization near 25~T corresponds to a phase transition that is accompanied by an electrical phase transition.

A more detailed analysis of the magnetization, capacitance, and electric polarization data, aimed at further elucidating the multiferroic and magnetoelectric coupling of this crystal, will be presented in a forthcoming publication \cite{won2025multi}.

\subsection{Comparison with commercial equipment}
\begin{figure}
\includegraphics[width=1\columnwidth]{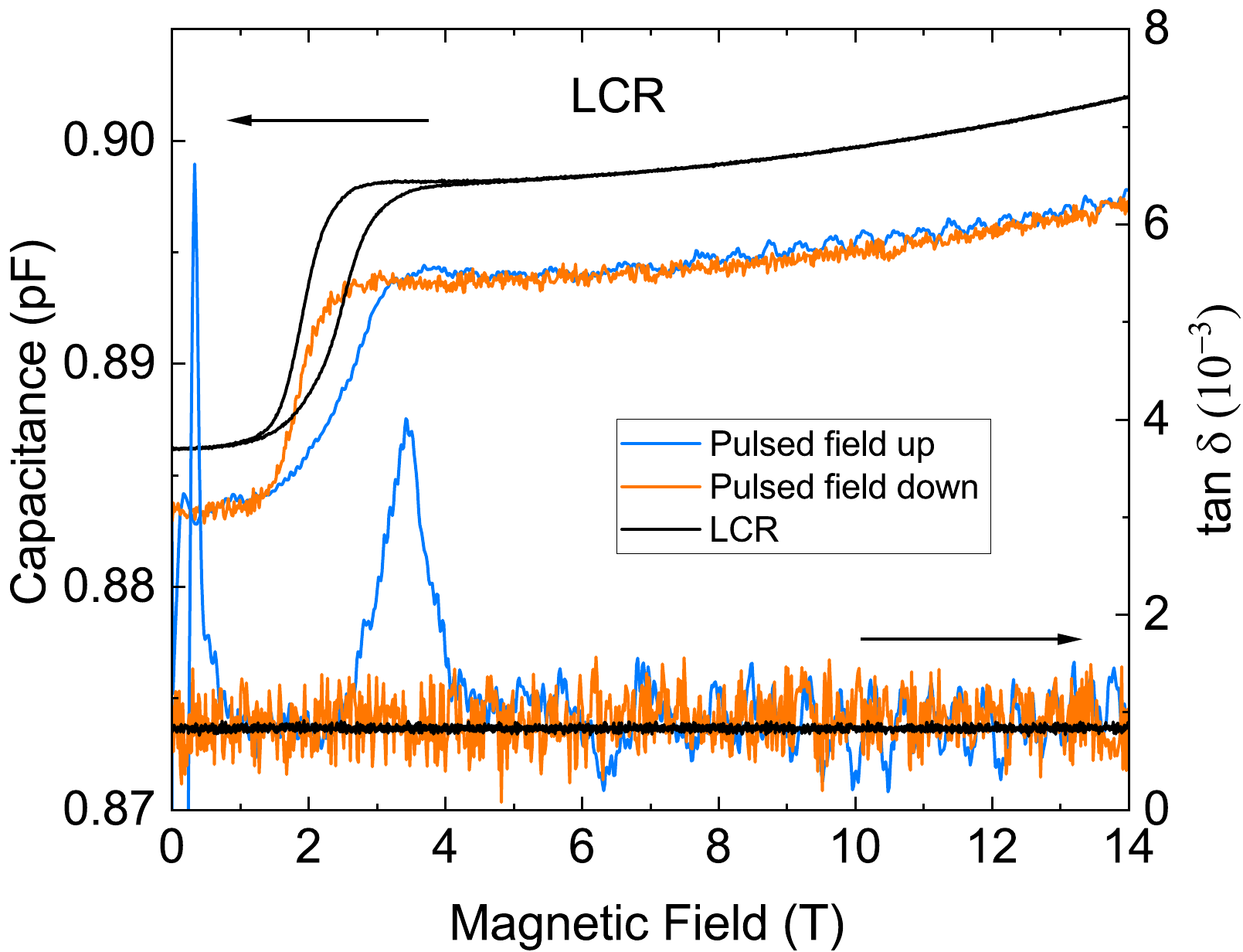}
\caption{\label{fig:LCt} (a) Capacitance and tan($\delta$) measured by a commercially available LCR meter in DC field and capacitance and tan($\delta$) measured in pulsed field with the same magnetic field and electric voltage configuration.}
\end{figure}
Fig.~\ref{fig:LCt} presents a comparison plot illustrating the differences between the capacitance measurement technique we developed and a commercially available LCR meter (Keysight E4980A). The frequency, excitation voltage, and averaging time were 101 kHz, 1 V, and 6, respectively. While our technique measured capacitance values approximately 0.003~pF lower than those obtained with the LCR meter which could be due to the residual inductance and resistance of the coaxial cable that is not preventable in this measurement set up, the overall shape of the capacitance curve remained highly consistent between the two methods. Furthermore, the variation of capacitance as a function of the magnetic field showed excellent agreement between our technique and the LCR meter, confirming the reliability of our approach. The noise level in our measurements was determined to be approximately 0.001~pF. We anticipate that this noise can be reduced by an order of magnitude through data averaging, particularly by  increasing the number of recorded data points. This suggests that with further optimization, our technique could achieve even greater precision in capacitance measurements.

For tan($\delta$), the overall values obtained using both the LCR meter and our measurement technique with the noise level is about 1$\times 10^{-3}$ were nearly identical. However, while no anomalies were observed in a DC magnetic field (with a maximum field sweep rate of 0.01~T/sec), a pronounced peak emerged around 3.5~T during pulsed field measurements, where the maximum sweep rate reached approximately 15~kT/sec. If the observed hysteresis loop originates from changes in magnetic and electric domains, it is plausible that under a rapidly varying magnetic field, these domains undergo rapid reconfiguration to achieve equilibrium, resulting in significant energy dissipation. This suggests that the dissipation mechanism is highly sensitive to the field sweep rate, further supporting the strong coupling between magnetic and electric domain dynamics in this system.

\section{Conclusion and Outlook}
In this work, we have developed and implemented a capacitance measurement technique optimized for use in pulsed magnetic fields, demonstrating its capability to capture subtle magnetic and electric phase transitions. Our method shows excellent agreement with commercial LCR meters in DC field conditions while revealing unique features under rapidly changing pulsed field environments. Specifically, we observed a pronounced hysteresis in capacitance that aligns with magnetization measurements, confirming a strong magnetoelectric coupling in NiCo$_{2}$TeO$_{6}$. Additionally, the presence of a significant peak in tan($\delta$) under fast magnetic field sweeps suggests an energy dissipation process likely tied to rapid domain reconfiguration. These results highlight the advantage of our capacitance technique in detecting field-induced phase transitions that might otherwise remain elusive in conventional magnetization measurements.

Looking ahead, this capacitance measurement approach opens new avenues for investigating multiferroic and magnetoelectric materials in rapidly changing conditions. Additionally, leveraging this technique to develop other measurement methods—such as capacitive torque magnetometry, magnetostriction measurements, and magnetic-field-independent thermometry for highly sensitive magnetocaloric and heat capacity measurements in pulsed fields—could provide a more comprehensive understanding of magnetoelectric interactions in complex materials. The ability to detect subtle electronic and magnetic state changes in pulsed fields highlights the versatility of this method, making it a powerful tool for future high-field research. Its potential applications extend beyond magnetoelectrics to a wide range of correlated electron systems, topological materials, and quantum phase transitions, offering new insights into emergent phenomena under diverse conditions.

\section{Data availability}
The data that support the findings of this study are available from the corresponding author upon reasonable request.

\section*{Acknowledgments} \label{sec:acknowledgements}
We thank Fedor Balakirev, Christopher A. Mizzi, and Gabriel Silva Freitas for useful discussions. This work from W. K. P., S. Z., S. L., V. S. Z., M. L. was performed at the National High Magnetic Field Laboratory, which is supported by National Science Foundation Cooperative Agreement No. DMR-2128556* and the State of Florida and the U.S. Department of Energy. W. K. P. acknowledges Seaborg Institute. M. L. and V. S. Z. acknowledge LDRD and Institute for Materials Science at LANL for their support. The work at Rutgers University was supported by the DOE under Grant No. DOE: DE-FG02-07ER46382. The work at Pohang University of Science and Technology was supported by the National Research Foundation of Korea (NRF) funded by the Ministry of Science and ICT (No. RS-2022-NR068223).

\bibliography{main}

\end{document}